\documentclass[pra,aps,twocolumn,showpacs]{revtex4}
\usepackage{amsmath,graphicx,epsfig,euscript}

\def\adot{\dot{\alpha}}
\def\adota{\dot{\alpha}/\alpha}

\def\cm{\mbox{~cm}}
\def\invcm{\mbox{~cm$^{-1}$}}

\def\mHz{\mbox{~mHz}}
\def\Hz{\mbox{~Hz}}
\def\kHz{\mbox{~kHz}}
\def\MHz{\mbox{~MHz}}
\def\GHz{\mbox{~GHz}}

\def\muG{\mbox{~$\mu$G}}

\def\G{\mbox{~G}}

\def\invs{\mbox{~s$^{-1}$}}

\def\mV{\mbox{~mV}}
\def\K{\mbox{~K}}

\def\fig_width{3. in} % width of single column figure in PR
\draft

\newlength{\defbaselineskip}
\newcommand{\setlinespacing}[1]%
           {\setlength{\baselineskip}{#1 \defbaselineskip}}

\begin{document}

\title{Towards a sensitive search for variation of the fine structure constant using radio-frequency E1 transitions in atomic dysprosium}
%%%----------------------------------------------------------------------
\author{A. T. Nguyen}
\email{atn@socrates.berkeley.edu} \affiliation{Department of
Physics, University of California at Berkeley, Berkeley,
California 94720-7300}
\author{D. Budker}
\email{budker@socrates.berkeley.edu} \affiliation{Department of Physics, University of
California at Berkeley, Berkeley, California 94720-7300} \affiliation{Nuclear Science
Division, Lawrence Berkeley National Laboratory, Berkeley, California 94720}
\author{S. K. Lamoreaux}
\email{lamore@lanl.gov}
\author{J. R. Torgerson}
\email{torgerson@lanl.gov} \affiliation{University of California,
Los Alamos National Laboratory, Physics Division, P-23, MS-H803,
Los Alamos, New Mexico 87545}

\date{\today}

%%%----------------------------------------------------------------------

%\doublespacing

\begin{abstract}
It has been proposed that the radio-frequency electric-dipole (E1)
transition between two nearly degenerate opposite-parity states in
atomic dysprosium should be highly sensitive to possible temporal
variation of the fine structure constant ($\alpha$) [V. A. Dzuba,
V. V. Flambaum, and J. K. Webb, Phys. Rev. A {\bf 59}, 230
(1999)]. We analyze here an experimental realization of the
proposed search in progress in our laboratory, which involves
monitoring the E1 transition frequency over a period of time using
direct frequency counting techniques. We estimate that a
statistical sensitivity of $|\adota| \sim 10^{-18}$/yr may be
achieved and discuss possible systematic effects that may limit
such a measurement.
\end{abstract}
\pacs{PACS. 06.20.Jr, 32.30.Bv}

%%%----------------------------------------------------------------------
%06.20.Jr   Determination of fundamental constants

%32.30.Bv   Radio-frequency, microwave, and infrared spectra
%%%----------------------------------------------------------------------

%\doublespacing

\maketitle

\section{Introduction}

Variation of the fundamental constants of nature would signify new
physics beyond the Standard Model, as discussed in a recent review
\cite{uzan03}. Various theories constructed to unify gravity with
the other forces allow or necessitate such a variation
\cite{marciano84,barrow87,damour94,damour02}. Of recent interest
is the astrophysical evidence for a variation of the fine
structure constant $\alpha$. From an analysis of quasar absorption
spectra \cite{webb01} over the redshift range $.5<z<3.5$, a
$4\sigma$ deviation of $\Delta\alpha/\alpha = (-0.72\pm
0.18)\times 10^{-5}$ from zero was reported. Although these data
hint at nonlinear dependence of $\alpha$ with time, for simplicity
we assume a linear shift over $\sim 10^{10}$ years, which
corresponds to a temporal variation of $\adota = (7.2 \pm
1.8)\times 10^{-16}$/yr. The current best terrestrial limit, over
a much shorter time scale of 2 billion years, is $|\adota| <
10^{-18}$/yr \cite{shlyakhter76,damour96,fujii00}, which comes
from an analysis of geophysical data obtained from a natural
fission reactor at Oklo (Gabon) which operated $1.8\times10^9$
years ago. Observational measurements like these, however, are
subject to numerous assumptions which tend to complicate the
interpretation. For example, questions have been recently raised
regarding the reliability of the Oklo analysis \cite{lamoreaux}.

Laboratory searches have numerous advantages, but have, thus far,
placed weaker limits on $\adot$. For example, a comparison between
H-maser and Hg$^+$ gave $\adota \leq 3.7 \times 10^{-14}$/yr
\cite{prestage95} while a similar comparison between Rb and Cs
microwave clocks yielded $|\adota| \leq 1.6 \times 10^{-15}$/yr
\cite{marion03}. A limit of $|\adota| < 1.2\times 10^{-15}$/yr was
obtained from a comparison of a Hg$^+$ optical clock to a Cs
microwave clock \cite{bize03} using a frequency comb.

It has been suggested \cite{dzubaPRL99,dzubaPRA99} that the
electric-dipole (E1) transition between two nearly degenerate
opposite-parity states in atomic dysprosium (Dy; Z=66) should be
highly sensitive to variations in $\alpha$. Indeed, a recent
calculation \cite{dzuba03} supports this conclusion. An
experimental search utilizing these states is currently underway
and is discussed here. We provide an analysis of possible
systematic effects and show that this search could ultimately
reach a sensitivity of $|\adota| \sim 10^{-18}$/yr.

\section{Variation of Alpha in Dysprosium}

Tests of variation of $\alpha$ in atomic systems rely upon the
fact that relativistic corrections depend differently on $\alpha$
for different energy levels. The total energy of a level can be
written as
%---------------------------------------------------------------
\begin{align}
    E = E_0 + q(\alpha^2/\alpha_0^2 - 1)+ {\mathcal{O}}(\alpha^3),
\end{align}
%---------------------------------------------------------------
where $E_0$ is the present-day energy, $\alpha_0$ is the
present-day value of the fine structure constant, and $q$ is a
coefficient which determines the sensitivity to variations of
$\alpha$. This coefficient mainly depends upon the electronic
configuration of the level. A recent calculation \cite{dzuba03},
utilizing relativistic Hartree-Fock and configuration interaction
methods, found values of $q$ for two nearly degenerate
opposite-parity states in atomic dysprosium that are both large
and of opposite sign. For the even-parity state (designated as A),
$q_A/hc = 6008\invcm$ while, for the odd-parity state (designated
as B), $q_B/hc = -23708\invcm$. The time variation of the
transition frequency between levels A and B is (for $\alpha
\approx \alpha_0$)
%---------------------------------------------------------------
\begin{align}
    \dot\nu = 2\frac{q_B-q_A}{h}\adot/\alpha \sim -2 \times
    10^{15}\Hz~\adot/\alpha.
\end{align}
%---------------------------------------------------------------
In other words, $|\adota| = 10^{-15}$/yr, implies $|\dot{\nu}| =
2\Hz$/yr.

The statistical uncertainty of determining the central frequency
from a resonance lineshape is $\delta \nu \sim \gamma / \sqrt{N}$,
where $N$ is the number of counts and $\gamma$ is the transition
width. In the dysprosium system, $\gamma = 20\kHz$ is determined
mainly by the lifetime of state A, $\tau_A = 7.9~\mu$s ($\tau_B
> 200~\mu$s) \cite{budker94}. A reasonable counting rate of
$10^9\invs$, corresponds to a statistical sensitivity to $\nu_0$
of $\sim 0.6~\tau^{-1/2}\Hz\sqrt{\mbox{s}}$ where $\tau$ is the
integration time in seconds. After an integration time of one day,
$\delta \nu \sim 2\mHz$, thus allowing for a statistical
sensitivity of $|\adota| \sim 10^{-18}$/yr for two measurements
separated by a year's time.

It should be noted that the sensitivity to a variation in $\alpha$
is not proportional to $\delta\nu/\nu$. In fact, if $\nu$ is
sufficiently small, the uncertainty in its determination is no
longer limited by the relative uncertainty of the reference clock
frequency ($\nu_{clock}$), but rather by other experimental
uncertainties. For example, a modest Cs clock provides an absolute
accuracy of $10^{-12}$. In the Dy transitions considered here,
typical frequencies are $\sim 1\GHz$ or smaller. Thus the clock
uncertainty will only become an issue when other uncertainties can
be reduced to the level of
%---------------------------------------------------------------
\begin{align}
    \nu \frac{\delta \nu_{clock}}{\nu_{clock}}\sim 10^{-3}\Hz,
\end{align}
%---------------------------------------------------------------
and even then can be overcome in a straightforward way by using a
better reference clock.

The relatively modest requirement for the reference clock also
means that the Dy experiment is insensitive to a variation of the
clock's frequency due to a change in fundamental constants. For
example, if $\alpha$ varies by $10^{-15}$, the fractional change
in the Cs clock frequency will be $2 \times 10^{-15}$, as the
hyperfine transition frequency employed in the clock is
proportional to $\alpha^2$. This variation is negligible in
comparison with the clock stability of $10^{-12}$.

%Note that the effect of a variation of $\alpha$ on the clock
%frequency is also negligible. For example, if $\alpha$ varies by
%$10^{-15}$, the fractional change in the Cs clock frequency will
%be $2 \times 10^{-15}$, as the hyperfine transition frequency
%employed in the clock is proportional to $\alpha^2$. This
%variation is negligible in comparison with the required clock
%stability of $10^{-12}$.

\begin{widetext}

\begin{figure}
\centerline{\psfig{figure=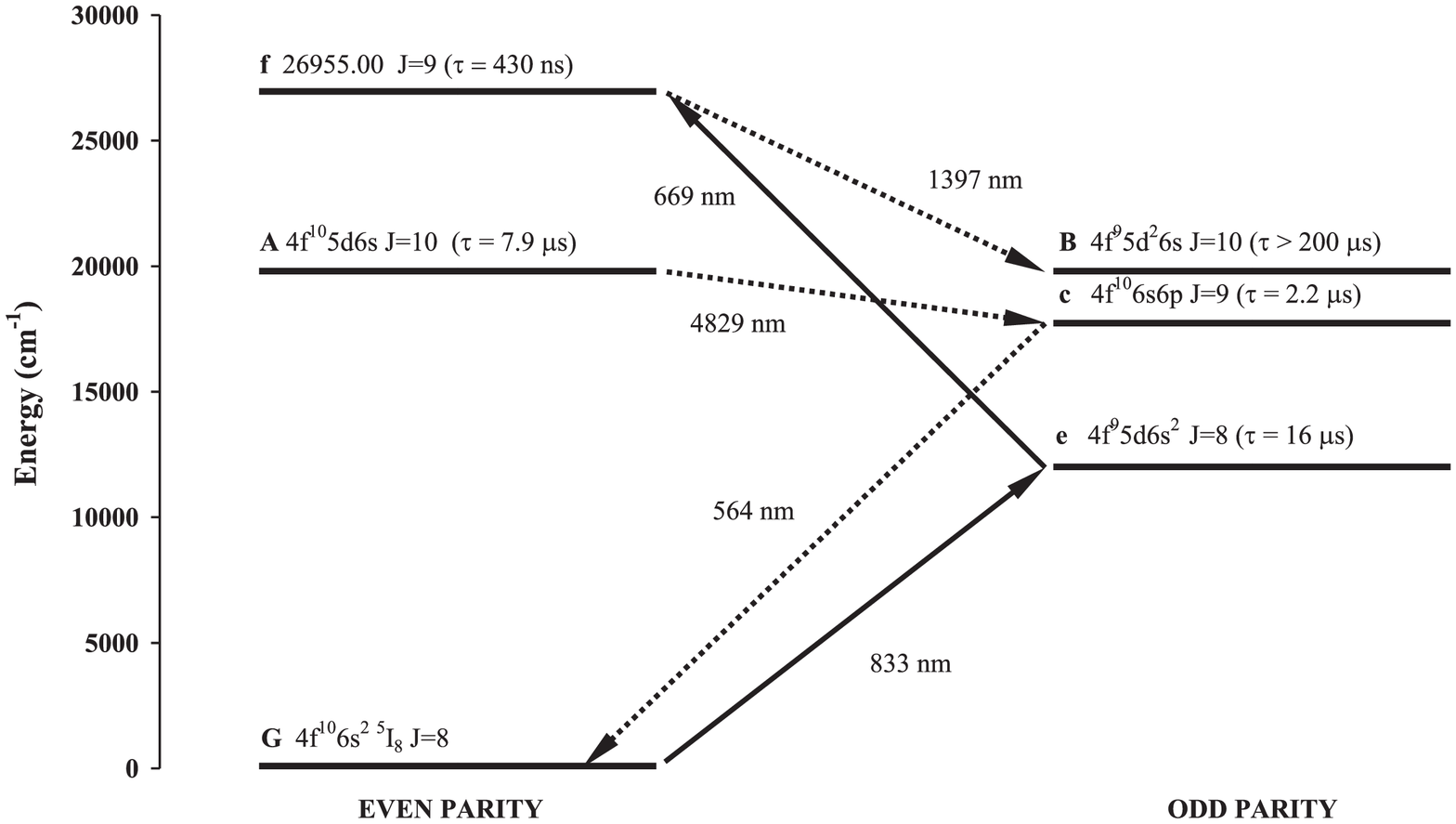,width=6 in}}
\caption{Partial energy-level diagram of dysprosium showing the
present scheme to populate level B and detect the population of
level A. Solid lines: absorption. Dashed lines: spontaneous
emission.} \label{elevels}
\end{figure}

\end{widetext}

\section{Experimental Technique}

\subsection{Overview}

The experimental search for a variation of the E1 transition
frequency between the two opposite-parity states will proceed as
follows. As shown in Fig.~\ref{elevels}, atoms are populated
\cite{nguyen00} to the longer-lived odd-parity state B ($\tau_B >
200~\mu$s \cite{budker94}) via three transitions. The first two
transitions require light at 833 nm and 669 nm, respectively. The
third transition involves spontaneous decay with a branching ratio
of 30\% and the emission of 1397-nm light. Once in state B, atoms
are transferred to state A with an rf electric field, whose
frequency is referenced to a commercial Cs frequency standard. A
resonance lineshape is attained by scanning the rf frequency and
monitoring 564-nm light from the second step of fluorescence from
state A.

\subsection{rf transitions}

\begin{figure}
\smallskip \centerline{\psfig{figure=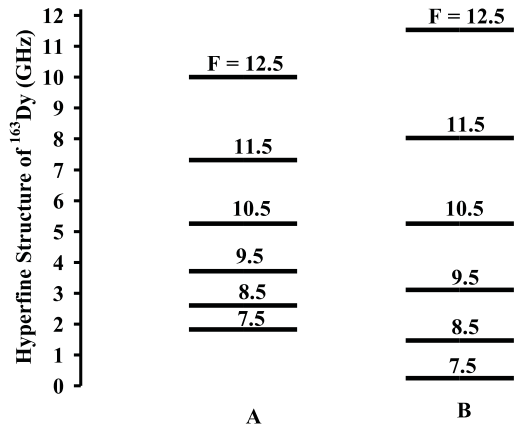,width=2.7 in}}
\caption{Hyperfine structure of of A and B levels of $^{163}$Dy.
Zero energy is chosen for the lowest hyperfine component.}
\label{163hfs}
\end{figure}

Because the energy separation between the nearly degenerate levels
is on the order of hyperfine splittings and isotope shift energies
(Fig.~\ref{163hfs}), and because dysprosium has seven stable
isotopes, there are many choices of rf frequencies.
Table~\ref{E1freq} shows rf transitions with frequencies below
$2\GHz$ calculated from measured hyperfine constants and isotope
shifts \cite{budker94}. The smallest transition frequency
($3.1\MHz$) occurs for the $F=10.5$ components of $^{163}$Dy (the
same transition is used in a search for parity nonconservation
(PNC) \cite{nguyen97}). A low-frequency transition that offers a
higher counting rate is the $235\MHz$ transition in $^{162}$Dy.
This is due to a large isotopic abundance and the fact that there
is no hyperfine splitting to dilute the atomic population. As
discussed below, the choice of rf transition is also important in
regard to sensitivity to systematic effects.

\begin{table}
\caption{Calculated E1 transition frequencies using the hyperfine
constants and isotope shifts from Ref. \cite{budker94}. $||d_F||$
and $||d_J||=1.5(1) \times 10^{-2}$ ea$_0$ $\approx 19\kHz/$(V/cm)
are reduced matrix elements. The last three columns list the
isotope mass numbers (and abundances) and total angular momenta of
the states involved in the transition.}
\medskip
%p{.75in} p{.75in} p{.75in} p{.5in} p{.5in}
\begin{tabular}{c c c c c}
    \hline \hline
    \rule{0ex}{4ex}
    $\nu_{rf}$          & ~~~$||d_F||/||d_J||$~~~     & ~~Mass No.      & ~~~~~~F$_A$~~~     & ~~~F$_B$~~~     \\
    ~~(\text{MHz})~~      &                           & ~~(Abund.)      &           &           \\
    \hline
    -1328.6             & 1.00                      & ~~160           & ~~10        & 10        \\
                        &                           & ~~(2\%)         &           &           \\
$\begin{matrix}
    -1856.4 \\
    -1714.7 \\
    -1249.7 \\
    -962.3  \\
    -791.5  \\
    -349.2  \\
    -172.7  \\
    68.9    \\
    514.0   \\
    1096.9  \\
    1576.0  \\
\end{matrix}$     &
$\left.\begin{matrix}
    0.15    \\
    1.04    \\
    0.99    \\
    0.15    \\
    0.94    \\
    0.89    \\
    0.19    \\
    0.86    \\
    0.20    \\
    0.19    \\
    0.15    \\
\end{matrix}\right.$          &
$\left.\begin{matrix}
            \\
            \\
            \\
            \\
            \\
            \\
            \\
            \\
            \\
            \\
\end{matrix}\right\}$$\begin{matrix} 161 \\ (19\%) \end{matrix}$          &
~~$\begin{matrix}
    7.5     \\
    11.5    \\
    10.5    \\
    12.5    \\
    9.5     \\
    8.5     \\
    11.5    \\
    7.5     \\
    10.5    \\
    9.5     \\
    8.5     \\
\end{matrix}$    &
$\begin{matrix}
    8.5     \\
    11.5    \\
    10.5    \\
    11.5    \\
    9.5     \\
    8.5     \\
    10.5    \\
    7.5     \\
    9.5     \\
    8.5     \\
    7.5     \\
\end{matrix}$\\
                        &                           &                 &           &           \\
    -234.7              & 1.00                      & ~~162             & ~~10        & 10        \\
                        &                           & ~~(26\%)        &           &           \\
$\begin{matrix}
    -1967.8 \\
    -1581.3 \\
    -1134.9 \\
    -609.7  \\
    -363.2  \\
    3.1     \\
    504.6   \\
    713.1   \\
    1531.0  \\
    1543.9  \\
\end{matrix}$     &
$\left.\begin{matrix}
    0.15    \\
    0.86    \\
    0.89    \\
    0.94    \\
    0.15    \\
    0.99    \\
    0.19    \\
    1.04    \\
    1.10    \\
    0.20    \\
\end{matrix}\right.$          &
$\left.\begin{matrix}
            \\
            \\
            \\
            \\
            \\
            \\
            \\
            \\
            \\
\end{matrix}\right\}$$\begin{matrix} 163 \\ (25\%) \end{matrix}$          &
~~$\begin{matrix}
    12.5    \\
    7.5     \\
    8.5     \\
    9.5     \\
    7.5     \\
    10.5    \\
    8.5     \\
    11.5    \\
    12.5    \\
    9.5     \\
\end{matrix}$    &
$\begin{matrix}
    11.5    \\
    7.5     \\
    8.5     \\
    9.5     \\
    8.5     \\
    10.5    \\
    9.5     \\
    11.5    \\
    12.5    \\
    10.5    \\
\end{matrix}$\\
                        &                           &                 &           &           \\
    753.5               & 1.00                      & ~~164             & ~~10        & 10        \\
                        &                           & ~~(28\%)        &           &           \\
    \hline \hline
\end{tabular}
\label{E1freq}
\end{table}

\subsection{Apparatus}

We have studied states A and B extensively as a system to measure
atomic PNC effects \cite{nguyen97}. Although our current apparatus
is optimized for a PNC experiment, it is suitable for a
measurement of $\adot$ with only minor modifications. We describe
this system here in order to make a realistic evaluation of a
possible experiment.

\begin{figure}
\centerline{\psfig{figure=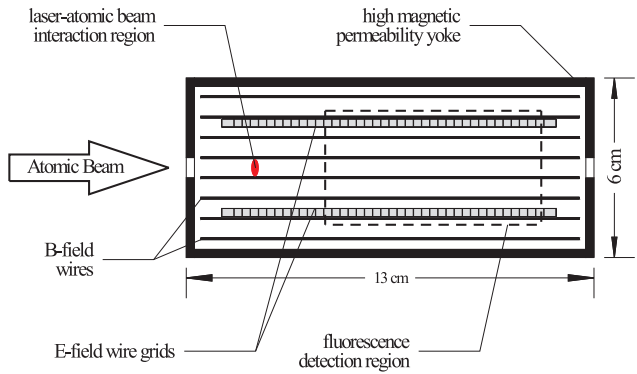,width=3.5 in}}
\caption{Side view of interaction region currently optimized for
PNC search.} \label{intregion}
\end{figure}

An atomic beam is produced by an effusive oven operating at
$1500$~K . The atoms pass through collimators and approach the
interaction region (Fig.~\ref{intregion}) where the desired
electric and magnetic fields are produced. Inside the interaction
region, the atoms encounter the laser beams used in the population
scheme. The rf electric field is formed between two wire grids
which are used in order to minimize surface area and thus stray
charge accumulation. Typical applied electric-field amplitudes are
$\sim 5$ V/cm. If necessary, a magnetic field of up to $\sim 4\G$
can be applied parallel to the electric field. It is produced by
eight turns of gold-plated copper wire surrounding the interaction
region. The entire interaction region is placed inside a magnetic
shield. As atoms decay from state A, 564-nm light from the second
step of fluorescence (Fig.~\ref{elevels}) is collected by a light
pipe and detected by a photomultiplier tube. The extent of the
light pipe area is shown by the dashed box in
Fig.~\ref{intregion}.

\begin{figure}
\centerline{\psfig{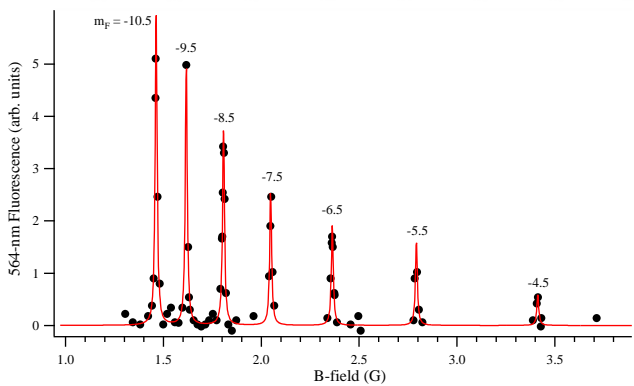}}
\caption{Zeeman-level-crossing resonances for $^{163}$Dy. State B
is populated as shown in Fig.~\ref{elevels} in the presence of an
electric field of $0.4$ V/cm. The population of state A, resulting
from electric field-induced mixing of states A and B, is monitored
by observing 564-nm fluorescence (Fig.~\ref{elevels}). The solid
line represents the best curve fit using the appropriate initial
level separation \cite{budker94} and g-values \cite{NBS}.}
\label{zeeman}
\end{figure}

An example of a resonance lineshape attained with this apparatus
is shown in Fig.~\ref{zeeman}. Here, a magnetic field is scanned
in the presence of a dc electric field revealing
Zeeman-level-crossing resonances for $^{163}$Dy. For the
measurement of $\adot$, the signal-to-noise will be much greater
as individual Zeeman resonances will not be resolved.

The interaction region can be improved by making it shorter, as
most atoms decay within twice the lifetime of state A ($\tau_A =
7.9~\mu$s). Given a mean atomic velocity of $5\times 10^4$ cm/s,
an appropriate length will be $\sim 8~$mm. Furthermore, a shorter
interaction region will allow for improved light collection
efficiency, better suppression of background oven light, and
better control over E- and B-fields.

\subsection{Lineshape}

The natural lineshape is a Lorentzian with a width ($\gamma
\approx 20\kHz$) determined mainly by the lifetime of state A
($\tau_A = 7.9~\mu$s). To a first approximation, both transit-time
and Doppler broadening change the lineshape in a symmetrical
fashion. For an interaction length $\sim 10\cm$ and a mean atomic
velocity of $v = 5\times 10^4$ cm/s, the transit-time width is
$\sim 5\kHz$. The Doppler shift (for an atom moving along the
propagation direction of a free electromagnetic wave of frequency
$\nu$) goes as $\sim \nu(v/c)$, which for a $1\GHz$ transition is
$\sim 2\kHz$. Thus, both transit-time and Doppler broadening
contributions to the width are much smaller than the natural
linewidth. Moreover, note that symmetrical broadening of the
resonance does not change the central frequency. In addition,
although an asymmetrical resonance can cause an apparent shift of
the central frequency, this is not important for the search for
variation of $\alpha$ as long as the asymmetry does not change
between measurements.

In the following section, we discuss mechanisms leading to
possible lineshape asymmetry and estimate the corresponding
shifts.

\section{Systematic Effects}

High statistical sensitivity implies that our technique will
likely be limited by how well we can control systematic effects.
Below we analyze possible systematics and give estimates of how
well various parameters of the experiment need to be controlled in
order to achieve a sensitivity of $|\adota| \sim 10^{-18}$/yr. A
summary of systematic effects and their estimated sizes is given
in Table~\ref{systematics}.

\begin{table}
\begin{center}
\caption{Estimated sizes of systematic effects thus far
considered.} \label{systematics}
\medskip
\begin{tabular}{p{2.4in} p{.9in}}
    \hline \hline
    \rule{0ex}{4ex}
    Systematic Shifts & Estimated Size ($\Hz$) \\ \hline
        ac Stark      $^{\dag}$                     & $\sim (0.1 - 30)$             \\
        Doppler effect                              & $< 0.2$                       \\
        room temp. black-body radiation             & $\lesssim  0.1$               \\
        oven black-body radiation                   & $\lesssim 0.02$               \\
        dc Stark$^{\dag}$                           & $\sim (10^{-4} - 10^{-2})$    \\
        collisional effects                         & $(1 - 10)\times 10^{-4}$      \\
        Millman effect                              & $\lesssim 5 \times 10^{-4}$   \\
        quadrupole moment                           & $\lesssim 10^{-5}$            \\
        Zeeman shift in stray B-field               & $\lesssim 10^{-5}$            \\
        \hline \hline
        $^{\dag}$transition dependent
\end{tabular}
\end{center}
\end{table}

\subsection{dc Stark and quadrupole shifts}

Variations of a stray dc electric field ($E_{dc}$) can give rise
to time-varying frequency shifts for the transition frequency
between states $m$ and $n$:
%---------------------------------------------------------------
\begin{align}
    \delta\nu_{dc} = \sum_{j \neq m} \frac{d_{mj}^2 E_{dc}^2}{\Delta_{mj}} - \sum_{k \neq n} \frac{d_{nk}^2 E_{dc}^2}{\Delta_{nk}},
    \label{dcStark}
\end{align}
%---------------------------------------------------------------
where $d_{mj}$ and $\Delta_{mj}=E_m-E_j$ are the dipole matrix
element and energy separation, respectively, between states $m$
and $j$. The sum in Eq.~(\ref{dcStark}) is taken over all states,
including hyperfine levels in the case of an odd isotope.

The reduced dipole matrix element for the B~$\rightarrow$~A
transition is $||d_J||=1.5(1) \times 10^{-2}$ ea$_0$ $\approx
19\kHz/$(V/cm)~\cite{budker94}. For large $J$, the maximum z
projection of the dipole moment $\approx ||d_J||/\sqrt{2J} =
4\kHz/$(V/cm). Matrix elements connecting all levels with the same
configurations as A and B have similar values~\cite{dzubapriv}.
Thus, between levels within $2000\invcm$ of the A and B levels,
dipole matrix elements are relatively small, which reduces the
sensitivity to systematics related to stray electric fields.

In an earlier PNC search, we reported stray dc E-fields,
presumably due to stray charge accumulation on the electrode
surfaces, that varied on time-scales from hours to months, and had
a typical magnitude of $50\mV/$cm~\cite{nguyen97}. Using this
value for the field and $d \sim 4\kHz/$(V/cm), we estimate the
shift for transition frequencies in the range $3-1000\MHz$ to be
$\sim 10^{-4}-10^{-2}\Hz$. The stray electric field can also be
measured at a few mV/cm level using the atoms
themselves~\cite{nguyen97} and cancelled by the application of an
external electric field.

An important systematic in Hg$^+$ optical clock
experiments~\cite{itano00} is the electric quadrupole shift due to
a stray-field gradient ($\nabla E$). Based upon the size of the
interaction region and the homogeneity of the electric field  for
the Dy beam experiment ($10^{-3}$)~\cite{nguyen97}, a conservative
estimate gives $\nabla E \sim 10 {\mbox{ mV/cm$^2$}}$. Assuming a
typical a quadrupole moment $Q \sim 1 {\mbox{ ea$_0^2$}}$ leads to
a quadrupole shift estimated to be $\lesssim 10^{-5}\Hz$ which is
negligible.

\subsection{ac Stark shift}

Fluctuations of the rf electric-field amplitude ($E$) can also
lead to time-varying shifts. The second-order ac Stark shift for a
two-level system is given by:
%---------------------------------------------------------------
\begin{align}
    \delta\nu_{ac} = \frac{d^2 E^2}{2}Re\left\{\frac{1}{\Delta-\nu+i\gamma} + \frac{1}{\Delta+\nu+i\gamma}\right\}, \label{acshift1}
\end{align}
%---------------------------------------------------------------
where $\nu$ is the applied rf frequency, $d$ is the dipole matrix
element, and $\Delta$ is the energy separation. Because the first
term is an odd function of detuning, only the second so-called
Bloch-Siegert term contributes to an actual shift of the central
frequency of the resonance lineshape. For $\nu = \Delta\gg \gamma$
and a transition near saturation for which $d^2 E^2/\gamma^2 \sim
1$, the corresponding shift is $\delta\nu_{ac} \sim
\gamma^2/(4\Delta)$. Hence for $\gamma = 20\kHz$ and transition
frequencies $\sim 3-1000\MHz$, the shift varies from $\sim
0.1-30\Hz$. For the $3.1\MHz$-transition, for which the shift is
largest, in order to achieve a sensitivity to frequency shifts of
a few mHz, the amplitude stability must be better than $10^{-4}$.
However, this requirement becomes much less stringent for higher
frequency transitions. For $\nu \sim 1\GHz$, only a modest control
at a level of a few percent is required.

We now consider frequency shifts due to all other levels on the
transition frequency between levels $m$ and $n$:
%---------------------------------------------------------------
\begin{align}\nonumber
    \delta\nu_{ac}' =& \sum_{j \neq m,n}\frac{d_{mj}^2
    E^2}{4}\left(\frac{1}{\Delta_{mj}-\nu}+\frac{1}{\Delta_{mj}+\nu}\right)\\
    &-\sum_{k \neq m,n}\frac{d_{nk}^2 E^2}{4}\left(\frac{1}{\Delta_{nk}-\nu}+\frac{1}{\Delta_{nk}+\nu}\right), \label{acshift2}
\end{align}
%---------------------------------------------------------------
where $d_{mj}$ is the dipole matrix element and $\Delta_{mj}=E_m -
E_j$ is the energy separation between levels $m$ and $j$. The
transition widths have been ignored because these levels are
off-resonant. For even isotopes, this shift is $\sim 0.3\Hz$ and
is mostly determined by levels which are  $> 2000\invcm$ away with
dipole matrix elements $\sim 1$ ea$_0$. A comparable shift occurs
for odd isotopes, assuming a particular choice of hyperfine
transitions for which $\Delta_{mj} - \nu \sim 1 \GHz$ and $d_{mj}
\sim 4\kHz/$(V/cm). Thus, taking into account the shifts due to
other levels does not lead to more stringent requirements on the
amplitude stability of the rf electric field beyond those obtained
in the two-level approximation.

\subsection{Stray magnetic fields}

The residual magnetic field ($B$) can be controlled to $\sim
1\muG$ in the magnetically shielded interaction region. This
corresponds to Zeeman shifts $g_F \mu_0 B \sim 1\Hz$, where
$\mu_0$ is the Bohr magneton and $g_F$ is the Land\'{e} g-factor
(Table~\ref{gvalues}). If the Zeeman sublevels are equally
populated but still unresolved, then this field does not shift the
central frequency of a resonance lineshape but rather only
broadens the lineshape with a corresponding width of $\sim F
(g_{FB}-g_{FA}) \mu_0 B \approx 1\Hz$. However, an imbalance of
sublevel populations can lead to asymmetric broadening of the
resonance lineshape, causing an apparent shift in the central
frequency. In the experimental geometry discussed
(Fig.~\ref{intregion}), such an imbalance may be caused by
residual circular polarization coupled with misalignments of the
propagation direction of the linearly polarized light beams used
in the first and second step of the population. Because residual
circular polarization can be controlled to the level of $\sim
10^{-5}$ using standard polarimetric techniques, the shift of the
resonance frequency can be made $\ll 10^{-5}\Hz$.

\begin{table}
\begin{center}
\caption{g-values for the hyperfine levels of states A and B
(g$_{JA} = 1.21$ and g$_{JB} = 1.367$ \cite{NBS}).}
\label{gvalues}
\medskip
\begin{tabular}{p{.5in} p{.5in} p{.5in}}
    \hline \hline
    \rule{0ex}{4ex}
    F & g$_{FA}$ & g$_{FB}$ \\ \hline
        &      \\
        7.5     & 1.57 & 1.77 \\
        8.5     & 1.36 & 1.54 \\
        9.5     & 1.22 & 1.38 \\
        10.5    & 1.11 & 1.26 \\
        11.5    & 1.03 & 1.16 \\
        12.5    & 0.968  & 1.09 \\ \hline \hline

\end{tabular}
\end{center}
\end{table}

\subsection{Millman effect in electric resonance}

Misalignments of the atomic beam and the geometry of the
electrodes can cause the direction of the rf E-field to rotate, as
seen by a moving atom. This may lead to frequency
shifts~\cite{grabner50} analogous to the Millman effect
encountered in magnetic resonance~\cite{millman39}. Let $\Omega$
be the frequency of this apparent E-field rotation. Because an
oscillating field can be decomposed into two counter-rotating
components, one of these components is shifted by $+\Omega$ and
the other by $-\Omega$.

\begin{figure}
\centerline{\psfig{figure=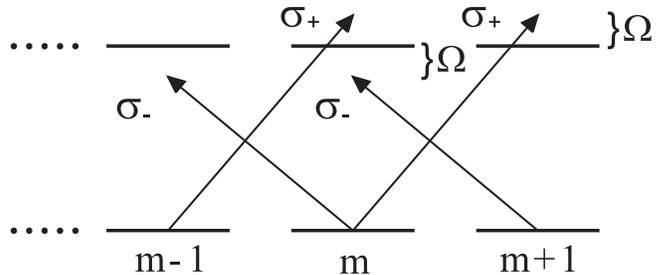,width=3.5 in}}
\caption{Electric field frequency shifts due to the Millman
effect.} \label{millman}
\end{figure}

Using the measured homogeneity of our electric field, we estimate
$\Omega \lesssim 50\Hz$. However, for a resonance lineshape with
unresolved sublevels, the central frequency remains largely
unaffected. This is not true in the case of magnetic resonance
(where transitions occur between magnetically split sublevels of
the same level) and can be explained as follows: In the basis in
which the quantization axis is perpendicular to the electric
field, there are only $\sigma_+$ and $\sigma_-$ transitions. The
Millman effect causes shifts in the frequencies of these
transitions as illustrated in Fig.~\ref{millman} for a $\Delta J =
0$ transition on resonance (where we have assumed that the
apparent electric field rotation is in the same sense as the
$\sigma_+$ component). If the sublevels of the initial state are
equally populated and $\Omega \ll \gamma$, then these shifts only
lead to a broadening of the lineshape. However, if there is an
imbalance of sublevel populations the broadening is asymmetric,
which can cause an apparent shift in the central frequency by an
amount $\sim \xi \Omega$, where $\xi$ is the degree of atomic
orientation. As mentioned earlier, residual circular polarization
in the linearly polarized light, used in the first and second step
of the population, may induce atomic orientation, but can be
controlled to the level of $\sim 10^{-5}$, which is sufficient for
the desired sensitivity to $\alpha$. Furthermore, it should be
noted that it is not the magnitude of the frequency shift but
rather the stability that is important.

\subsection{Black-body radiation}

Black-body radiation (BBR) can cause ac Stark
shifts~\cite{palchikov02}. The rms value of the black-body
radiation electric field is
%---------------------------------------------------------------
\begin{align}
    \langle E^2 \rangle = (8.3\mbox{
    V/cm})^2[T(\mbox{K})/300\K]^4.
\end{align}
%---------------------------------------------------------------
A shift in the transition frequency arises from the difference in
the ac Stark shifts experienced by levels $A$ and $B$. We can give
a rough estimate of this Stark shift ($\delta \nu_{BB}$) for one
of these levels using Eq.~(\ref{acshift2}):
%---------------------------------------------------------------
\begin{align}
    \delta\nu_{BB} \sim \frac{d^2 \langle E^2
    \rangle}{4(\Delta-\nu_{BB})},
\end{align}
%---------------------------------------------------------------
where $\Delta$ is a characteristic atomic energy scale, $\nu_{BB}$
is a characteristic frequency of room-temperature BBR, and $d = 1$
ea$_0$ is a typical optical transition dipole moment. Due to
cancellations of contributions from nearby energy levels above and
below the A and B levels, we assume that the net shift comes
mainly from levels with large energy separations: $\nu_{BB} \ll
\Delta \sim 10^{14}\Hz$, gives an estimate of $\delta \nu_{BB}
\lesssim 0.1\Hz$.

The interaction region is illuminated by much hotter BBR from the
atomic oven (T$\sim 1500\K$) which is $\sim 15\cm$ away. However,
the effect is smaller due to the decreased solid angle:
%---------------------------------------------------------------
\begin{eqnarray}\nonumber
    \delta\nu_{oven} &\sim& \delta\nu_{BB} [T(\mbox{K})/300\K]^4 \left(\frac{\Delta
    \Omega}{4\pi}\right),\\ \nonumber
    &=& 0.1\Hz \left(\frac{1500\K}{300\K}\right)^4 \left(\frac{1\cm^2}{4\pi
    ~15^2\cm^2}\right),\\
    &=& 0.02\Hz,
\end{eqnarray}
%---------------------------------------------------------------
where we have assumed that an atom in the interaction region sees
$1\cm^2$ of hot surface.

The BBR shifts are large compared to the desired level of
sensitivity and are difficult to eliminate. However, these shifts
can be kept constant to the desired level by stabilizing the
temperature of the interaction region (to $\sim 2~^\circ$C) and
the oven (to $\sim 30~^\circ$C). For the oven, possible changes in
surface emissivity will be a concern and requires further
investigation.

\subsection{Collisional shifts}

Because the typical cross-sections of the atomic collisions are on
the order of $10^{-14}\ $cm${^2}$, most atoms in the atomic beam
do not experience collisions with other atoms before they
encounter the back wall of the interaction region. Collisional
shifts of $\sim 1-10$ MHz/Torr (corresponding to the
above-mentioned collisional cross-section values) are typically
found for atomic transitions when the width is measured as a
function of the atomic pressure. A possible effect of the rare
collisions in the beam can be estimated by extrapolating these
numbers to the low pressure values corresponding to the beam
environment. Assuming a $10\ $MHz/Torr shift due to collisions
with residual gas, in order to keep time-dependent shifts below a
few mHz, the residual gas pressure must be stable at a level $\sim
10^{-10}$ Torr. This can be achieved with standard
ultra-high-vacuum equipment. The pressure stability and
composition of the residual gas will be monitored with a residual
gas analyzer.

Note that we may find that collisional shifts are, in fact, much
smaller than the above estimate because the A-B transition is
nominally between f and d inner-shell electrons~\cite{vedenin85}.
Another important effect that may reduce the effect of collisions
on the measurement of the variation of $\alpha$ is that it is
likely that an atom experiencing a collision will be quenched to a
different atomic state, and, thus, will not produce a 564-nm
fluorescent photon, and will therefore avoid detection.

A similar consideration applies to collisions between Dy atoms
themselves. In the absence of collisional quenching, the
worst-case estimate is that the present Dy density ($\sim 10^{10}$
atoms/cm$^3$) could give a shift of up to $\sim 2\Hz$. This effect
will be investigated by measuring transition frequencies as a
function of the atomic beam intensity and, if needed, the Dy beam
intensity can be monitored and stabilized during measurement.
Measuring a transition frequency at several different Dy-beam
intensities will allow extrapolation to the collision-free value
of the frequency. Technically, it is possible to vary the density
of atomic in the atomic beam without significantly affecting other
conditions of the experiment. For example, black-body radiation
intensity in the interaction region can be maintained constant by
separately controlling the front and back temperatures of the Dy
oven. The BBR intensity is determined by the temperature of the
front part of the oven near the nozzle, while the flux of atoms is
determined by the lower temperature towards the back of the oven.

Another concern is a possible collisional wall-shift resulting
from a fraction of the atoms reentering the interaction region
upon reflection from its back wall. This effect is suppressed by a
number of small factors: reflection probability, solid angle for
entering the interaction region, the probability for an atom to
avoid quenching in the wall collision, etc. If necessary,
cryo-cooling of the interaction-region walls may be employed to
reduce the probability for an atom to bounce from it.

%Collisional shifts are typically $\sim 1-10$ MHz/Torr for atomic
%transitions. In the worst case scenario, to keep shifts below a
%few mHz, the residual gas pressure must be stable at a level $\sim
%10^{-10}$ Torr. This can be achieved with standard UHV equipment.
%The pressure stability and composition of the residual gas will be
%monitored with a residual gas analyzer. However, we may find that
%collisional shifts are much smaller because the A-B transition is
%nominally between f and d inner-shell electrons~\cite{vedenin85}.
%In addition, quenching cross-sections, which effectively remove
%atoms from the beam, are expected to be large and so will reduce
%the effect of collisional shifts.
%
%Another concern is collisional shifts due to Dy atoms themselves.
%In the absence of collisional quenching, the worst-case estimate
%is that the present Dy density ($\sim 10^{10}$ atoms/cm$^3$) gives
%a shift of $\sim 2\Hz$. This effect will be investigated and, if
%needed, the Dy beam intensity can be monitored and stabilized.

\subsection{Doppler Shift}

The effect of the first-order Doppler shift is estimated to be
small. To see this, we model the electric field
(Fig.~\ref{intregion}) as a standing wave constructed from two
traveling waves counter-propagating co-linearly with the direction
of the atomic beam. Due to ohmic losses inside the conductors, the
amplitude of each wave gets attenuated. A difference in the
intensity of these waves at the location of the atoms leads to
asymmetric Doppler broadening of the lineshape, and thus an
apparent shift. We estimate this intensity difference by first
considering the B-field induced by the time-varying E-field. This
B-field penetrates inside the metal to within a skin-depth,
inducing currents from which the power loss can be readily
calculated~\cite{jackson}. This simple model gives a fractional
power difference $\propto \nu^{5/2}$. Since the Doppler shift for
a given traveling-wave power is $\propto \nu$, the shift is
$\propto \nu^{7/2}$. For a transition frequency of $1\GHz$, we
estimate the asymmetry to be $\lesssim 10^{-4}$ and the
corresponding shift is $\sim 0.2\Hz$. One can imagine a factor of
$\sim 10$ suppression if rf power is fed to the plates in a
symmetric fashion. Thus, it is only required that this shift be
stable to $\sim 10\%$ in order to achieve a sensitivity of a few
$\mHz$.

In addition to the first-order Doppler shift, we also consider the
second-order effect. Depending upon the rf transition, the second
order Doppler shift is $\sim 10^{-3}-10^{-6}\Hz$, which is
sufficiently small.

\subsection{Techniques to control systematics}

\begin{figure}
\centerline{\psfig{figure=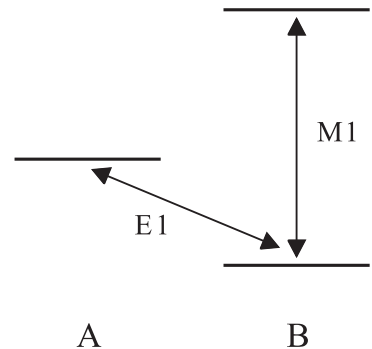,width=2.0in}}
\caption{Simultaneous measurement of an M1 transition frequency
for a nonzero nuclear spin isotope facilitates the control of
systematics.} \label{e1m1}
\end{figure}

\begin{figure}
\centerline{\psfig{figure=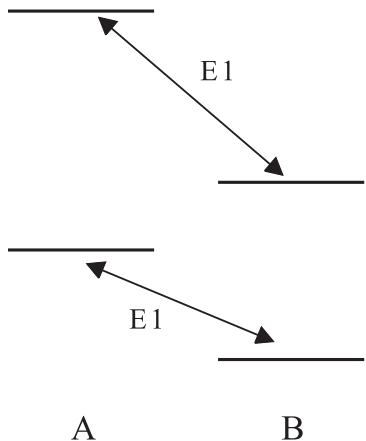,width=2.0in}}
\caption{Simultaneous measurement of two (or more) E1 transition
frequencies to facilitate the control of systematics. This method
works with the two zero-spin isotopes, $^{162}$Dy and $^{164}$Dy.}
\label{e1e1}
\end{figure}

A powerful method to detect and eliminate possible sources of
systematic shifts common to both levels is to simultaneously
measure the transition frequency between hyperfine levels of a
given parity state. The reason is that, because the levels
involved have the same relativistic corrections, this frequency is
insensitive to variations of $\alpha$. One possible scheme is to
excite an M1 transition, e.~g., as shown in Fig.~\ref{e1m1}, whose
frequency can be monitored by looking for disappearance in the
fluorescence from level $A$ for a fixed E1 transition frequency.
Alternatively, we can utilize another E1 transition as shown in
Fig.~\ref{e1e1}. In this scenario, the effect $\alpha$ variation
is twice as large in the sum of the two frequencies, while the
difference is insensitive to $\alpha$ variations.

Furthermore, one can compare E1 transitions for the two abundant
isotopes with zero nuclear spin ($^{162}$Dy and $^{164}$Dy). The
counting rate is significantly higher and the level structure is
much simpler without hyperfine interactions.

\section{Discussion}

In summary, rf E1 transitions in Dy provide an attractive system
in which to test the temporal variation of $\alpha$. The
frequencies of these transitions can be directly counted. For a
limit of $\adota < 10^{-15}$/yr, the shift was calculated to be
$\approx 2\Hz$/yr. At present, a statistical sensitivity of
$0.6\Hz$ in one second of integration time is achievable.
Knowledge of systematic effects is critical to this experiment.
Preliminary analysis shows that it may be possible to control them
at a level corresponding to $|\adota| \sim 10^{-18}$/yr, a level
of sensitivity that would rival that of the most stringent
observational limit set by the Oklo natural reactor.

\acknowledgments

We thank D. F. Kimball, M. G. Kozlov, J. E. Stalnaker, and V.  V.
Yashchuk for valuable discussions. This work was supported in part
by the UC Berkeley-LANL CLE program and a NIST Precision
Measurement Grant. D.B. also acknowledges the support of the
Miller Institute for Basic Research in Science.

\end{document}